# On ponderomotive metallization of magnetospheric plasma

A.V. Guglielmi[1], A.S. Potapov[2], F.Z. Feygin[1]

[1] *Schmidt Institute of Physics of the Earth, Russian Academy of Sciences, Moscow, Russia, guglielmi@mail.ru*

[2] *Institute of Solar-Terrestrial Physics of Siberian Branch of Russian Academy of Sciences, Irkutsk, Russia, potapov@iszf.irk.ru*

**Abstract.** The problem of ponderomotive separation of ions with different charge-to-mass ratios under the influence of Alfvén waves, which permanently exist in the magnetosphere in the form of geomagnetic pulsations, is posed. Formulas are derived for partial ponderomotive forces acting on light and heavy (metal) ions. In the quasi-hydrodynamic approximation, it was found that the Clarke number, which characterizes the metallicity of the plasma, is maximum at the minimum of the magnetic field on the field line along which the Alfvén wave propagates.

## 1. Introduction

In the linear approximation, the propagation of a monochromatic electromagnetic wave in a plasma is accompanied by the harmonic motion of electrons and ions. If we move to a quadratic approximation in terms of the wave amplitude and average over the period of harmonic oscillations, it turns out that a ponderomotive force **f** acts on a unit volume of plasma. Let us assume that the plasma is in an external magnetic field **B**, as is the case in the Earth's magnetosphere, and we will decompose the force into longitudinal and transverse components with respect to **B**. The general phenomenological expression for **f** is given in the fundamental monograph [1]. We will focus on the longitudinal component

$$f_\| = \frac{1}{16\pi}\left[(\varepsilon_{\alpha\beta} - \delta_{\alpha\beta})\nabla_\| E_\alpha^* E_\beta + E_\alpha^* E_\beta \frac{\partial \varepsilon_{\alpha\beta}}{\partial \mathbf{B}} \partial \mathbf{B}\right], \quad (1)$$

since under its influence the plasma is accelerated along geomagnetic field lines, while the transverse component only leads to plasma drift at a constant speed. Here $\varepsilon_{\alpha\beta}(\omega)$ is the plasma dielectric constant tensor [2, 3], **E** is the amplitude of electric field oscillations, and $\omega$ is the wave frequency.

In the Earth's magnetosphere, the most powerful wave activity occurs in the magnetohydrodynamic (MHD) range. We will limit ourselves to the analysis of the ponderomotive force of Alfvén waves [4], which are represented in the magnetosphere by a rich variety of geomagnetic pulsations [5–7]. It was previously shown [8–11] that under the influence of Alfvén waves and ion-cyclotron waves, a noticeable redistribution of plasma occurs along geomagnetic field lines. It was found that the ponderomotive force "rakes" the plasma towards the minimum of the magnetic field, i.e. towards the plane of the geomagnetic equator with a dipole approximation of the external magnetic field. As a result, an equatorial compaction of the plasma occurs.

From a theoretical point of view, the traveling Alfvén wave is of particular interest. The wave trajectory coincides with the geomagnetic field line, and at each point of the trajectory the relation

$$\nabla_\| E_\perp^2 = E_\perp^2 \nabla_\| \ln \frac{B^2}{\sqrt{\rho}} \quad (2)$$

is satisfied where $\rho = \sum m_i N_i$ is the plasma density, $m_i$ and $N_i$ are the mass and concentration of ions of the $i$ type, and the summation is carried out over all types of ions. From (1) taking into account (2) a simple expression for the ponderomotive force follows:

$$f_\| = -\frac{1}{8}\left(\frac{cE_\perp}{B}\right)^2 \nabla_\| \rho, \tag{3}$$

that is convenient for calculations.

Strictly speaking, formula (3) is applicable when the strong inequality $\omega \ll \min\{\Omega_i\}$ is satisfied. Here $\Omega_i = e_i B / m_i c$ is the gyrofrequency of an ion with charge $e$, $c$ is the speed of light. Under this condition, all ions move with the same acceleration under the action of the Alfvén wave. This strictly limits the range of problems that can be solved analytically. In particular, it does not allow analyzing the ponderomotive separation of ions with different charge-to-mass ratios.

However, the condition $\omega \ll \min\{\Omega_i\}$ can be weakened in the special case of hydrogen plasma containing only a small admixture of massive ions. For example, the magnetosphere is filled predominantly with protons of solar origin with a small admixture of oxygen ions of terrestrial (ionospheric) origin. Mass spectrometry measurements on satellites indicate the presence of small impurities of other elements, including iron, but for simplicity and clarity we will talk about a two-component plasma containing $H^+$ and $O^+$ ions. The Alfvén wave retains dispersion and polarization properties in the frequency range $\omega \ll \Omega_{H^+}$ with the additional limitation that the wave frequency does not fall into a narrow frequency band in a very small vicinity of the gyrofrequency of oxygen ions $\Omega_{O^+}$.

These two restrictions, without being too strict, leave us with the opportunity to show that the ponderomotive forces acting on ions with different charge-to-mass ratios are greater the smaller the specified ratio. As a result, ponderomotive separation of different types of ions occurs and the chemical composition of the plasma changes. In particular, heavy ions accumulate in places where the plasma is compacted, i.e. in the vicinity of the geomagnetic equator in the dipole model of the magnetosphere. Let us recall that in cosmophysics it is customary to call all chemical elements heavier than hydrogen and helium metals, so we can talk about the metallization of magnetospheric plasma in the magnetic field minima on the field lines along which Alfvén waves propagate.

## 2. Partial forces

Let us decompose the ponderomotive force (1) into the sum of partial ponderomotive forces: $f_{\|} = \sum f_{\|s}$. Here, $s = e, i$ and the subscript $e$ denotes electrons. Let's take into account the relation

$$\varepsilon_{\alpha\beta} = \delta_{\alpha\beta} + \frac{4\pi i}{\omega} \sigma_{\alpha\beta} \tag{4}$$

and use the additivity of the complex electrical conductivity tensor: $\sigma_{\alpha\beta} = \sum \sigma_{\alpha\beta s}$. After this we will express $f_{\|s}$ through $\sigma_{\alpha\beta s}$. Here $\sigma_{\alpha\beta s}$ is the contribution of charged particles of the $s$ type to electrical conductivity.

Let us select the geomagnetic field line along which the Alfvén wave propagates and introduce an accompanying coordinate system $(x, y, z)$ in the vicinity of the line so that the $x$ axis is directed along the binormal and the $z$ axis along the tangent. Let the Alfvén wave have polarization $\mathbf{E} = (E_\perp, 0, 0)$. In accordance with formula (1), we need the following component of the dielectric constant tensor:

$$\varepsilon_{xx} = 1 + \sum \frac{\omega_{0s}^2}{\Omega_s^2 - \omega^2}. \tag{5}$$

Here $\omega_{0s} = \sqrt{4\pi e^2 N_s / m_s}$ is the plasma frequency, and $e$ is the elementary electric charge. For simplicity, we consider all ions to be singly charged. Using (1), (2), (5), we obtain

$$f_{\|s} = -\frac{E_\perp^2}{8\pi} \left( \frac{\omega_{0s}^2}{\Omega_s^2 - \omega^2} \right) \left[ \partial \ln \rho^{1/4} + \left( \frac{\omega^2}{\Omega_s^2 - \omega^2} \right) \partial \ln B \right], \tag{6}$$

where $\partial \equiv \partial/\partial z$.

Formula (6) is applicable, in particular, to hydrogen plasma with a small admixture of oxygen ions if two conditions are met. One of them is obvious: $\omega \ll \Omega_{H^+}$. To formulate the second condition, we use the so-called clarke number $\kappa = \rho_{O^+}/\rho_{H^+}$, where $\rho_{O^+}$ and $\rho_{H^+}$ are the densities of oxygen and hydrogen ions, respectively. (The term "clark number", introduced by A.E. Fersman, is used in geophysics and cosmophysics in a similar context.) The second condition for the applicability of formula (6) is that the value $|1-\omega/\Omega_{O^+}|$ must exceed the clarke number $\kappa \ll 1$. Taking these restrictions into account, the partial ponderomotive forces are equal to

$$f_{\|e} = -\frac{1}{8}\left(\frac{cE_\perp}{B}\right)^2 \rho_e \partial \ln \rho, \tag{7}$$

$$f_{\|H^+} = -\frac{1}{8}\left(\frac{cE_\perp}{B}\right)^2 \rho_{H^+} \partial \ln \rho, \tag{8}$$

$$f_{\|O^+} = -\frac{E_\perp^2}{8\pi}\left(\frac{\omega_{0O^+}^2}{\Omega_{O^+}^2 - \omega^2}\right)\left[\partial \ln \rho^{1/4} + \left(\frac{\omega^2}{\Omega_{O^+}^2 - \omega^2}\right)\partial \ln B\right]. \tag{9}$$

**3. Diffusion equilibrium**

Let us consider the static equilibrium of a two-component isothermal plasma in the quasi-hydrodynamic approximation. The balance of forces acting along the geomagnetic field line is described by the equations

$$\partial p_e = \rho_e g_\| - eNE_\| + f_{\|e}, \tag{10}$$

$$\partial p_i = \rho_i g_\| - eN_i E_\| + f_{\|i}. \tag{11}$$

Here $p_e = NT$ and $p_i = N_i T$ are the partial pressures of electrons and ions, $T$ is the temperature, $N$ is the electron concentration, $g_\|$ is the longitudinal projection of gravitational acceleration, $E_\parallel$ is the electric field of ambipolar diffusion. The index $i$ takes values 1 and 2 for quantities related to light and heavy ions, respectively. From (7) – (11), taking into account the quasineutrality condition $N = N_1 + N_2$, the expression for the ambipolar electric field follows

$$E_\parallel = -\frac{m_+}{2e}\left(g_\parallel + a_\parallel\right), \tag{12}$$

where $m_+ = \rho/N$ is the average ion mass and $a_\parallel = \left(f_{\parallel 1} + f_{\parallel 2}\right)/\rho$ is the ponderomotive acceleration. Let us substitute (12) into (11) and obtain two first-order nonlinear differential equations

$$T\partial N_i = \left(m_i - \frac{m_+}{2}\right)N_i g_\parallel - \frac{N_i}{2N}\left(f_{\parallel 1} + f_{\parallel 2}\right) + f_{\parallel i}, \tag{13}$$

describing the distribution of ions along the geomagnetic field line.

Taking into account the cumbersome formula (6), equations (13) can be studied in detail using numerical methods. We will limit ourselves to a qualitative analysis of plasma metallization in the vicinity of the minimum magnetic field on the field line along which the Alfvén wave propagates. In a dipole magnetosphere, the minimum is located at the point of intersection of the field line with the equatorial plane. The magnetic field increases quadratically with distance from the specified plane. Omitting calculations, we indicate the expression for the clarke number in a fairly narrow equatorial zone

$$\kappa(z) = \kappa(0)\exp\left(-\alpha z^2\right), \tag{14}$$

where

$$\alpha = \frac{e^2}{4m_2 T}\left(\frac{\omega E_\perp}{\Omega_2^2 - \omega^2}\right)^2 \frac{1}{B}\frac{d^3 B}{dz^2}. \tag{15}$$

Here the distance is measured along the field line from the point of its intersection with the equatorial plane. The quantities $\Omega_2$, $E_\perp$, $B$, and $d^3 B/dz^2$ refer to the point $z = 0$.

We see that $\kappa$ is maximum at the minimum of the geomagnetic field. The quantity $\kappa$ characterizing the metallicity of the plasma is proportional to the square of the wave amplitude and the smaller the frequency difference $\Omega_2 - \omega$, the greater it is.

## 4. Discussion

The geomagnetic field has a rather complex configuration [6]. In the direction towards the Sun, it is limited by the magnetopause, the position of which varies depending on the dynamic pressure of the solar wind. The magnetopause is approximately located at a distance of 50 thousand km from the Earth. In the antisolar direction, the field forms a long geomagnetic tail extending over hundreds of thousands of kilometers.

The dipole approximation of the field approximately corresponds to satellite observations at altitudes not exceeding approximately 30 thousand km. With further distance from the Earth in the direction of the Sun, a kind of bifurcation occurs: the equatorial minimum of the geomagnetic field splits into two minima located symmetrically relative to the equatorial plane [11]. In each of them, an increase in the clarke number is expected. In places where the field lines passing through the minima touch the magnetopause, two deep magnetic wells called cusps are formed. In the cusps, significant plasma metallization is expected due to heavy ions of terrestrial and solar origin.

Outside the magnetosphere, the trajectories of Alfvén waves coincide with the lines of force of the interplanetary magnetic field (IMF), the sources of which are located on the Sun. The IMF has a sector structure, with field minima located at the boundaries between sectors [12, 13]. In the interplanetary medium there is a permanent flow of Alfvén waves of very large amplitude propagating from the Sun [14]. We can confidently expect that the sector boundaries are highly metallized due to the action of the ponderomotive ion separation mechanism. It should, however, be emphasized that the supersonic solar wind blows in the interplanetary medium, so the theory of static equilibrium is not applicable in this case. The analysis of ponderomotive separation of ions in a non-uniform stationary flow is expected to be carried out in a separate work.

To conclude the discussion, let us point out the characteristic clarke number of oxygen in the magnetosphere. Instead of the weight clarke $\kappa = \rho_{O^+}/\rho_{H^+}$, we will indicate here the atomic Clarke $\xi = N_{O^+}/N_{H^+}$, since we use for evaluation the results of satellite measurements $N_{H^+}$ and $N_{O^+}$, which we borrow from the monograph [15]. In the daytime hemisphere of the magnetosphere in the equatorial plane at a distance of five Earth radii from the center of the Earth, the characteristic value of the oxygen clarke is equal to $\xi = 0.01$. It should be emphasized that the value is highly variable in time and space (see, for example, [16–18]).

## 5. Conclusion

We considered the issue of ponderomotive separation of ions with different charge-to-mass ratios under the influence of Alfvén waves and derived formulas for partial ponderomotive forces acting on light and heavy (metallic) ions. In the quasi-hydrodynamic approximation, a system of equations is obtained that describes the distribution of ions along magnetic field lines in the Earth's magnetosphere. It was found that the clarke number, which characterizes the metallicity of the plasma, is maximum at the minimum of the magnetic field on the field line along which the Alfvén wave propagates. The possibility of applying the theory to the analysis of mass spectrometric changes in the chemical composition of space plasma is indicated.

We express our gratitude to B.I. Klain for his interest in the work and valuable comments. This work was carried out with financial support from the Russian Foundation for Basic Research in the framework of project No. 19-05-